# Community detection in complex networks via node similarity, graph representation learning, and hierarchical clustering

Łukasz Brzozowski [a],*, Grzegorz Siudem [b], Marek Gagolewski [a,c]

[a] Warsaw University of Technology, Faculty of Mathematics and Information Science, ul. Koszykowa 75, 00-662 Warsaw, Poland
[b] Warsaw University of Technology, Faculty of Physics, ul. Koszykowa 75, 00-662 Warsaw, Poland
[c] Systems Research Institute, Polish Academy of Sciences, ul. Newelska 6, 01-447 Warsaw, Poland

ABSTRACT

Community detection is a significant challenge in analysing real graphs, including social, citation, biological, and many other networks. This article proposes three new, general, hierarchical frameworks to deal with this task. The introduced approach supports various linkage-based clustering algorithms, vertex proximity matrices, and graph representation learning models. We compare 364 module combinations on Stochastic Block Model graphs and real-life datasets. We observe that our best pipelines (Wasserman–Faust and the mutual information-based PPMI proximity, as well as the deep learning-based DNGR representations) perform competitively with the state-of-the-art Leiden and Louvain algorithms. At the same time, unlike the latter, they remain hierarchical. Thus, they output a series of nested partitions of all possible cardinalities which are compatible with each other. This feature is crucial when the number of correct partitions is unknown in advance, and we show that it is obtained at near-zero practical cost in terms of partition quality. On graphs composed of many small communities, our pipelines are more accurate than the reference solutions. Nevertheless, the downside of our approach is its time complexity, which confines the framework to networks of some $10^4$ vertices on personal machines.

## 1. Introduction

### 1.1. Background

More and more structures become available in the form of graphs and networks. The problem of community detection therein is a long-standing task in data science: a plethora of solutions have already been developed (see, e.g., [1–3]). Still, the topic seems far from exhausted [4]. What constitutes a community can be understood very broadly [1]. Therefore, many relatively different tools will aim to provide us with their own solutions to this task: is a community a group of tightly connected vertices, or is it a group of vertices that play similar roles in the graph's structure? Further, the sole existence of some mathematically definable relation allows us to construct a network in a given environment. However, such networks may be, for example, structured, hierarchical, temporal, evolving, or attributed. The various generalisations of the concept of a graph call for different community detection solutions tailored to each specific context [1,5,6].

Researchers in pattern recognition and statistical/machine learning have long known a similar problem under a different name: clustering [7]. At some level, we can say that the only difference between clustering and community detection is the data space considered: the former is defined on graphs whilst the latter is studied on other metric spaces, such as the Euclidean one [7]. This similarity resulted in some domain-specific approaches stemming from a common theoretical basis. For example, hierarchical algorithms may be successfully employed to solve both community detection and clustering tasks with little change in the implementation. However, with the need for more specialised methods [2,8], the approaches developed in the two aforementioned fields began to diverge.

### 1.2. Research objectives

Taking into account the growing popularity of graph machine learning (e.g., [9,10]), which includes graph representation learning specifically, we wish to help bridge the between-domain gap by exploring new ways of connecting community detection and clustering solutions.

Specifically, in this paper, we develop a framework for applying hierarchical clustering algorithms (e.g., [11,12]) to the task of community detection. The hierarchical approach is time-tested and turns out to be robust and easily applicable to graphs, because it relies solely on similarity metrics between pairs of nodes. As the problem of choosing the right measures for our task is crucial, we perform an extensive comparative analysis of twenty-four vertex similarity modelling techniques falling into the three following classes:

1. distance matrices obtained by considering various node similarity measures: the Wasserman–Faust measure [13], Neighbourhood overlap, i.e., the Jaccard coefficient of two neighbourhoods [1], $\ell$-step walk probability [1], Katz index [14], Rooted PageRank [15], a reciprocal-degree index related to that of Adamic and Adar [16], Blondel–Gajardo measure [17], and Positive Pointwise Mutual Information [18],
2. spectral versions of the above, i.e., eigendecompositions of the corresponding vertex similarity matrices,
3. distances on automatically discovered spaces using the representation learning algorithms: DeepWalk [19], DNGR [18], Graph Factorisation [20], GraRep [21], HARP [22], HOPE [15], Laplacian Eigenmaps [23], and Node2Vec [24].

* Corresponding author.
*E-mail addresses:* lukasz.brzozowski@pw.edu.pl (Łu. Brzozowski), grzegorz.siudem@pw.edu.pl (G. Siudem), marek.gagolewski@pw.edu.pl (M. Gagolewski).
*URLs:* http://if.pw.edu.pl/~siudem (G. Siudem), https://www.gagolewski.com/ (M. Gagolewski).

Combined with seven clustering settings, this gives 364 configurations, which we evaluate on both synthetic and real-life networks. We study how the tuning of the underlying parameters affects the quality of clustering and, in each case, determine the best linkage function.

We should emphasise that hierarchical approaches come with a set of advantages. Firstly, the communities are nested: depending on the selected number of clusters, we can observe how a large community decomposes into smaller ones, so the dendrogram is itself a description of the network's organisation rather than a by-product of obtaining one partition. Where a network is genuinely hierarchical, e.g., when it represents departments within divisions or topics within fields, that structure is available without deciding in advance at which level to look. Secondly, hierarchical approaches allow us to select any number of clusters, which is a feature often lacking in other methods. Leiden [25] and Louvain [26] are multi-level algorithms and do yield a few nested partitions, but at resolutions the algorithm arrives at rather than ones the user requests; there is in general no level with exactly $K$ communities for an arbitrary $K$.

Overall, we perform a comprehensive analysis of hierarchical clustering algorithms applied to the task of community detection. Our contributions are the following.

1. A modular framework in which hierarchical clustering algorithms are applied to graphs through any of three routes – node dissimilarities, the eigenvectors of a vertex similarity matrix, or a learned Euclidean representation – together with a rule for selecting the number of communities from the dendrogram, so that the method takes a graph and returns a partition (Section 3).
2. A comparative benchmark of 364 component configurations on 2880 stochastic block model graphs and six real-world networks, to our knowledge the most extensive comparison of vertex proximity models for hierarchical community detection to date. It is conducted under a protocol that separates the data used to select components from the data used to evaluate them, and reports the failure rate of every component (Sections 4 and 5).
3. A measurement of what the hierarchy costs: cutting the dendrogram by an internal criterion recovers at least 97% of the accuracy available when the true number of communities is supplied (Section 6.2).
4. Identification of the regime in which the framework is more accurate than Leiden and Louvain, and of the mechanism behind it. When a graph contains many small communities, modularity maximisation merges them; we measure that deficit and show that our advantage grows and shrinks with it. The effect reproduces on real networks (Sections 6.2 and 6.3).
5. A characterisation of the cost: run time grows as $n^{2.8}$ to $n^{2.9}$ against $n^{0.9}$ for Leiden, which places the practical limit of the approach at some $10^4$ vertices (Section 6.4).
6. A negative result: replacing the Laplacian with an arbitrary vertex similarity matrix does not yield a working spectral method, and for two of the measures we show the substitution to be disadvantageous (Section 6.5).

The paper is set out as follows. In Section 2, we introduce the node similarity measures and graph representation learning methods that serve as graph models in the task at hand, and position the framework against existing hierarchical approaches. In Section 3, we present the framework itself, including the selection of the number of communities. Section 4 describes the benchmark data, the evaluation protocol and the statistical methodology. In Section 5, we screen the components and select nine pipelines, which are then, namely in Section 6, carefully evaluated (including on real-world data). Finally, Section 7 discusses our findings and concludes the article.

## 2. Background and related work

In this section, we introduce the vertex proximity measures and the graph representation learning methods that serve as graph models in our framework. Then, we position this framework with respect to the existing hierarchical approaches to community detection.

### 2.1. Community detection

A common way of conducting scientific research is to find generalisations of the existing concepts to explain them better in connection with the definitions from other fields. In that sense, Network Science is not that different: why should we define a "small" graph, such as a citation network of machine learning papers, when we could define a more general citation network of all academic research? This approach allows us to work on larger structures, which usually means that we obtain better statistical approximations of some sought values. Nevertheless, this also inevitably means that our network becomes less homogeneous. In fact, while the whole citation network contains all the domains, we should expect that its subgraph induced by the vertices representing machine learning papers is connected more densely than the whole network on average. Such an induced subgraph is called a *community* [1]. In general, communities can be found in many naturally occurring networks, e.g., they may be based on the shared interests of social media users in a social network [27], on patterns in the propagation of delays in railway networks [28], or on the similarity between protein functions in biological networks [29].

Therefore, to understand the behaviour of a network correctly, it is essential to detect whether it contains communities and, if so, find which nodes belong to which clusters. However, the above description of a community is far from strict. In fact, there is no universally accepted definition [1,4]. Here we would like to present some metrics that could help define the community.

**Definition 1.** Let $H = (V_H, E_H)$ be an induced subgraph of graph $G = (V_G, E_G)$ and let $n_H = |V_H|$. We say that $e$ is an *internal* edge of $H$ if it connects two vertices in $V_H$, and an *inter-cluster* edge of $H$ if exactly one of its ends lies in $V_H$. Write $n = |V_G|$. As the maximum number of edges in $H$ is equal to $\binom{n_H}{2} = \frac{n_H(n_H-1)}{2}$, we define the *intra-* and *inter-cluster* densities of $H$, respectively, as (see [1]):

$$\nu_{\text{int}}(H) = \frac{\text{the number of internal edges of } H}{\binom{n_H}{2}}, \tag{1}$$

$$\nu_{\text{ext}}(H) = \frac{\text{the number of inter-cluster edges of } H}{n_H(n - n_H)}. \tag{2}$$

Intuitively, a community, regarded as an induced subgraph of $G$, should have higher intra-cluster density and lower inter-cluster density than the average link density of the graph $G$. To detect communities, we could look for a partition that either maximises the average value of $\nu_{\text{int}}$, minimises the average value of $\nu_{\text{ext}}$, or combines the two; for example, we could maximise the average value of $v_{\text{int}} - v_{\text{ext}}$ [30]. However, the problem with algorithms simply relying on the above-defined concepts of cluster density is that the posed tasks are often NP-complete. For example, finding all clusters in $G$ that have intra-cluster density $v_{\text{int}}$ higher than a given threshold $\zeta$ cannot be performed in polynomial time [31].

Another issue with the above definitions of cluster density is that we calculate them only relative to the observed network $G$. If $G$ is heterogeneous, it becomes difficult to precisely define the requirements for a subgraph to be a community. For example, we would require a community to have an extremely high density if it lies in a densely connected region of $G$, but we would accept a subgraph with a lower density in a sparse region in $G$. Such differences could lead to problems with proper task definition and unreliable results. On the other hand, we know that the Erdős–Rényi (ER) model [32] should not contain communities, as the edges are positioned randomly with equal probabilities. Based on that fact, Newman and Girvan introduced the concept of *modularity* [33].

**Definition 2.** Let $G = (V, E)$ and let us assume that we have partitioned $G$ into clusters and let $A$ be its adjacency matrix. Let $\tilde{G}$ be a *null model*, i.e., a graph that is structurally similar to $G$ but does not contain communities. Let $\tilde{A}_{i,j}$ be the expected number of edges between $i$ and $j$ in $\tilde{G}$ and let $C_i$ denote the cluster in $G$ containing the vertex $i$. Then, denoting the Kronecker delta with $\eth$, we define a partition quality function called *modularity* as:

$$Q = \frac{1}{2|E|} \sum_{i,j \in V} (A_{i,j} - \tilde{A}_{i,j})\, \eth(C_i, C_j). \tag{3}$$

Intuitively, given some partition of $G$, modularity compares the observed number of edges in the communities with the expected number of such edges in the null model. Various candidates for the null model have been investigated: the Erdős–Rényi model is the simplest of them all, but we could derive specific null models for particular classes of networks [34]. While explicitly maximising modularity is again NP-complete, it has become a popular evaluation metric for community detection algorithms [1].

Modularity is not, however, an unbiased objective. Fortunato and Barthélemy [35] showed that it suffers from a *resolution limit*: in a graph with $|E|$ edges, a modularity-maximising partition merges any two communities connected by at least one edge whenever each of them contains fewer than about $\sqrt{|E|}$ internal edges, however clearly separated the two may be. The threshold depends on the size of the whole graph rather than on the local structure, so the same pair of communities may be resolved in a small network and merged in a large one. Consequently, methods that optimise $Q$ systematically underestimate the number of communities in networks made up of many small groups. This is a structural property of modularity as an objective, rather than of any particular optimiser.

### 2.2. Node similarity

In the previous section, we have defined the community to be an induced subgraph, which is, in some sense, connected more strongly than its neighbourhood. However, we could look at the community detection problem more locally: a community could be a subgraph of *somewhat* similar vertices. In such an approach, we would like the communities to be subgraphs such that the intra-cluster edges connect similar vertices and the inter-cluster edges connect different ones [1]. Assuming that two vertices are similar if they are connected in the graph, we naturally return to the previous definition. However, the flexibility of similarity allows us to introduce other approaches to community detection. In this section, we will present some selected node similarity measures.

*Embedding-based measures [1].* Let us assume that we have embedded the graph $G$ in the space $\mathbb{R}^d$ for some $d$, i.e., that we have vector representations $X_v = (x_v^1, x_v^2, \dots, x_v^d) \in \mathbb{R}^d$ for each node $v$ of $G$. Regardless of how the embedding was obtained, we can now easily define the node similarity between two vertices $u$ and $v$ as a distance between $X_u$ and $X_v$, e.g., the Euclidean, Manhattan, or Chebyshev metric. Applying any non-increasing transformation turns such a distance into a similarity. We will discuss particular graph embedding methods in Section 2.3.

*Relationship-based measures [1,36].* Another class of similarity measures is based on the relationships encoded within the graph. The most straightforward function of the similarity between the vertices $u$ and $v$ of $G$ is $A_{u,v}$, i.e., the element of the graph's adjacency matrix indicating whether or not $u$ and $v$ are adjacent. However, we could also express their similarity in terms of the structure of their neighbourhoods, e.g., based on the Wasserman–Faust distance [13]:

$$d_{u,v}^{WF} = \sqrt{\sum_{k \neq u,v} (A_{u,k} - A_{v,k})^2}. \tag{4}$$

The Wasserman–Faust distance is one of the first measures based on the structural equivalence of the vertices. It does not rely on the adjacency of $u$ and $v$. A degree-weighted index related to that of Adamic and Adar [16] also incorporates the structural equivalence. We use the reciprocal-degree form:

$$S_{u,v}^{\text{RD}} = \sum_{w \in N(u) \cap N(v)} \frac{1}{2|N(w)|}, \tag{5}$$

rather than the logarithmic weighting $1/\log|N(w)|$ of the original index, since the reciprocal form admits the closed form $S^{\text{RD}} = A\hat{D}^{-1}A$ with $\hat{D} = \operatorname{diag}\big(\sum_j (A_{ij} + A_{ji})\big)$ and is therefore obtainable without enumerating common neighbourhoods. We refer to it as the reciprocal-degree index below. The intuition behind both formulae is that common neighbours having large neighbourhoods are less significant to the similarity between the vertices. Another similarity based on the neighbourhoods of $u$ and $v$ is the neighbourhood overlap, that is, the Jaccard coefficient of the two neighbourhoods:

$$S_{u,v}^{\text{NO}} = \frac{|N_G(u) \cap N_G(v)|}{|N_G(u) \cup N_G(v)|}. \tag{6}$$

*Walk-based measures.* In another approach, we could measure similarity between the vertices based on walks in the graph. Intuitively, if many paths connect two vertices, they would seem similar. Fortunately, the adjacency matrix of a graph captures all information about walks. It is known that for an unweighted graph $G$, the element $A_{u,v}^k$ counts the number of walks of length exactly $k$ between the vertices $u$ and $v$ [37]. For instance, the Katz index matrix [14] is:

$$S^{\text{Katz}} = \sum_{\ell=1}^{\infty} \beta^\ell A^\ell = \beta \cdot (I - \beta \cdot A)^{-1} \cdot A, \tag{7}$$

where $\beta$ is a decay parameter, and $I$ denotes the identity matrix. Then, the value $S_{u,v}^{\text{Katz}}$ gives the walk-based similarity between the vertices $u$ and $v$. The series converges only for $\beta < 1/\rho(A)$, where $\rho(A)$ is the spectral radius of the adjacency matrix; outside that range, the closed form on the right-hand side still evaluates, but it no longer represents a count of walks.

We may also be interested in whether the two vertices $u$ and $v$ often co-occur in the graph's random walks. This approach differs slightly from the previous one, as it assumes that the co-occurrence in random walks gives more practical information about the similarity between the two vertices than the Katz index. If $D$ is the degree matrix of $G$, then the probability that a random walk started at $u$ is located at $v$ after exactly $\ell$ steps is the corresponding entry of the $\ell$-th power of the transition matrix [38]:

$$p_{u,v}^{(\ell)} = \big((D^{-1}A)^\ell\big)_{u,v}. \tag{8}$$

We refer to Eq. (8) as the $\ell$-step transition probability.

Similarly, instead of using unbiased random walks, we could turn to more sophisticated tools like the Rooted PageRank algorithm [15,39]. It assumes that we "surf" through the vertices and walk to some neighbour with probability $\alpha$, or return to the starting vertex with probability $1-\alpha$. If $P = D^{-1}A$ is the transition probability matrix, we get that $S^{\mathrm{RPR}} = \alpha \cdot S^{\mathrm{RPR}} \cdot P + (1-\alpha) \cdot I$, which can be solved to obtain:

$$S^{\mathrm{RPR}} = (1-\alpha)(I - \alpha P)^{-1}. \tag{9}$$

The elements of this matrix correspond to node similarity based on the Rooted PageRank algorithm. Note that $I - \alpha P$ becomes singular as $\alpha \to 1$, so the restart probability must be kept strictly positive.

Finally, we can use random walks to estimate the value of Pointwise Mutual Information (PMI) [40] between the vertices. That is, we identify the similarity between $u$ and $v$ as:

$$\mathrm{PMI}(u,v) = \log \frac{p(u,v)}{p(u)p(v)}, \tag{10}$$

where $p(u)$ denotes the probability of the occurrence of $u$ in some random walk of a fixed length. Analogously, $p(u,v)$ denotes the probability that both vertices occur on the same random walk, calculated using powers of a transition matrix of a stochastic random walk on the graph. Usually, as the PMI values may be negative, we use the Positive Pointwise Mutual Information (PPMI), i.e., we replace all the negative elements with 0. We estimate the co-occurrence probabilities with the random-surfing procedure of Cao et al. [18], in which a walk continues to a neighbour with probability $\alpha$ and restarts at the source otherwise, accumulated over $T$ steps; $\alpha$ and $T$ are given in Section 4.2.

### 2.3. Graph representation learning

We will use several algorithms which can be loosely divided into four categories: direct optimisation-based, matrix factorisation-based, random walk-based, and autoencoder-based methods. Let us present a short description of each of the categories and their noteworthy instances. For the sake of clarity, throughout this section, we assume that we embed the nodes of a graph $G = (V, E)$ in $\mathbb{R}^d$, i.e., we obtain a matrix $Z \in \mathbb{R}^{|V| \times d}$ whose rows correspond to individual nodes. We also denote the adjacency matrix of $G$ with $A$, its Laplacian with $L$, and its degree matrix with $D$.

*Direct optimisation methods.* This class of methods generates the embedding of $G$ by solving a specific optimisation task. For example, in the Laplacian Eigenmaps [23] algorithm, we select $Z$ which minimises $\mathrm{tr}(Z^T L Z)$ subject to $Z^T D Z = I$, where $\mathrm{tr}$ is the trace of a matrix. Belkin and Niyogi [23] show that this optimisation task corresponds to a representation where adjacent nodes are possibly close in the resulting space while the non-adjacent nodes lie far from each other. Furthermore, in the Graph Factorisation (GF) [20] method, the similarity between the nodes $v_i$ and $v_j$ is approximated by the dot product of the corresponding representations, i.e., $s(v_i, v_j) \approx \langle Z_i, Z_j \rangle$.

*Matrix factorisation methods.* The second class of methods uses the Singular Value Decomposition (SVD), which creates a lower-dimensional approximation of higher-dimensional structures [41]. To obtain the embedding, we utilise the SVD on a matrix representation of a graph and again rely on the dot product $\langle Z_i, Z_j \rangle$ approximation of the similarity. However, neither of the two methods discussed below performs the SVD on the adjacency or the Laplacian matrix, on the grounds that these matrices do not represent node similarity accurately. GraRep [21] performs the decomposition on an $\ell$-step transition probability matrix, and HOPE [15] allows the use of the aforementioned Katz Index, Adamic–Adar, Rooted PageRank, or Jaccard similarity matrices. It must be noted that HOPE is a tool developed specifically for directed graphs with asymmetric similarity matrices.

*Random walk methods.* The next group of algorithms trains an embedding model based on random walks in the graph. They all use the Skip-Gram model [42], which is a tool developed for Natural Language Processing. This design decision can be justified by the similarity of node sequences in a random walk and the word sequences in a sentence. Here, DeepWalk [19] is the most basic but efficient solution, in which first-order Markov walks in a graph are fed to the neural network to generate the embedding. Node2Vec [24] was developed to improve DeepWalk by replacing the walks with biased second-order Markov walks. Let us note that Chen et al. [22] noticed inefficient weight propagation in the Skip-Gram [42] neural network for large graphs. Therefore, they developed a meta-algorithm where the Skip-Gram model is trained iteratively on a family of graphs of increasing sizes derived from $G$. The weights corresponding to the nodes are passed between the iterations, improving the final representation. The HARP algorithm may work with either DeepWalk or Node2Vec as the underlying model.

*Autoencoder methods.* These methods use an autoencoder neural network to generate an embedding. Specifically, a matrix representation of the graph $G$ is passed as input to both the neural network and the loss calculation. In that sense, autoencoder methods use a unary encoder in the described encoder–decoder architecture. Then, after the network is trained, the matrix $Z$ is retrieved from the middle of the autoencoder network. The two methods, SDNE [43] and DNGR [18], differ primarily in the input matrix representation, beyond slight architectural changes. SDNE uses the graph adjacency matrix $A$, and DNGR uses the Positive Pointwise Mutual Information matrix; we evaluate the latter.

A further class of methods, exemplified by Struc2Vec [44] and GraphWave [45], embeds vertices according to their *structural role* rather than their proximity in the graph, so that distant vertices with similar neighbourhoods are placed close together. Since the notion of community we adopt in Section 2 is a proximity-based one, these methods target a different task, and we do not evaluate them here.

### 2.4. Hierarchical approaches to community detection

The use of hierarchical methods has a long history in this field. The divisive algorithm of Girvan and Newman [29] removes edges of the highest betweenness one at a time and thereby produces a full dendrogram over the vertices; Clauset et al. [46] fit a hierarchical random graph model whose dendrogram is inferred rather than constructed greedily; and, more recently, Bonald et al. [47] obtain a dendrogram directly from a graph-adapted agglomerative rule.

The relationship between our framework and the modularity-based state of the art is as follows. Both Louvain [26] and Leiden [25] are *multi-level* algorithms: they alternate local moves with the aggregation of detected communities into super-nodes, and the sequence of aggregation passes does yield a small number of nested partitions. What they do not provide is control over the granularity. The number of communities is a by-product of maximising $Q$, the levels correspond to resolutions the algorithm arrives at rather than ones the user requests, and there is in general no level with exactly $K$ communities for an arbitrary $K$. A resolution parameter can be introduced and tuned to shift the outcome, but doing so requires re-running the algorithm and does not produce a hierarchy that is nested across the resulting partitions.

The framework we develop differs on precisely this axis. A single run produces a complete dendrogram, from which the partition into any number of communities between 1 and $|V|$ can be read off, and these partitions are nested by construction.

### 2.5. Representation learning for community detection

Representation learning has also been applied to community detection directly, rather than as a source of vertex proximities. vGraph [48] is a generative model that performs community detection and node embedding jointly, parametrising the community and node distributions with Euclidean representations; it permits overlapping communities, which makes it unsuitable where a partition is required. LCDEN [49] targets evolutionary networks, combining a Laplacian Eigenmaps loss with a deep sparse autoencoder so that the detected communities may follow the graph's evolution over time. The Structural Deep Clustering Network [50] couples a deep autoencoder with a Graph Convolutional Network [10] through a dual self-supervised module, and reports strong results on large attributed networks, outperforming solutions such as Variational Graph Autoencoders [51].

While the solutions mentioned here provide necessary context, we do not use them as references for comparison. SDCN and its relatives consume node attributes in addition to graph structure, whereas our framework, just like Leiden, Louvain, and the other reference algorithms we use, sees only the topology; vGraph solves the overlapping-community problem; LCDEN addresses temporal graphs. None of the three produces a hierarchy of nested partitions, which is the property under examination here. We refer the reader to the survey [2] for descriptions of other recent solutions, including deep learning models, and for a discussion of how young this line of research still is.

## 3. Distance-based graph clustering framework

### 3.1. Hierarchical clustering

Most hierarchical clustering methods are either divisive or agglomerative. DIANA [52] is an example of a divisive algorithm (see [53] for an application): starting from a single cluster $C_1$, we find the data point having maximal average dissimilarity to all the others and move it to a new cluster $C_2$. Then, all points that are now, on average, more similar to $C_2$ than $C_1$ are moved to $C_2$. The algorithm progresses iteratively until a predefined stopping requirement is met.

However, agglomerative algorithms are more popular due to better computational complexity and thus lie at the heart of our paper. Assuming a data set $S = \{s_1, s_2, \ldots, s_n\}$ equipped with a dissimilarity measure, we first partition $S$ such that each point belongs to its own cluster, i.e., we define a partition consisting of singletons, $C = \{C_1, C_2, \ldots, C_n\}$. Then, we merge pairs of clusters iteratively. The merging strategy is what differentiates various algorithms. Denoting by $d(s_i, s_j)$ the selected distance measure from $s_i$ to $s_j$, we merge the clusters $C_a$ and $C_b$ that minimise a selected linkage criterion $L(C_a, C_b)$, with the single (minimum), complete (maximum), average, and Ward (variance) linkages being the most common choices; see, e.g., [11]. Carried to completion, the procedure performs $n-1$ merges and thus induces a sequence of nested partitions $C^{(1)}, \ldots, C^{(n)}$, where $C^{(k)}$ consists of exactly $k$ clusters. This sequence, forming the dendrogram, is the object our framework works with.

It must be noted that hierarchical clustering methods may also be based on graphs. For example, we could generate a graph using $k$-nearest neighbours methods and define a criterion using the degrees of nodes [54].

The single-linkage criterion is relatively naïve and often does not yield suitable partitions. The remaining ones have higher time complexity. One way to bypass the complexity problem is to add additional requirements. For example, the Genie algorithm [12,55] uses a single-linkage criterion but simultaneously monitors the Gini index of the cluster size distribution, which is a measure of inequality. Once that index exceeds a predefined threshold, the next merge is restricted to involve a cluster of the smallest cardinality, which counteracts the tendency of single linkage to produce one dominant cluster alongside many small ones.

### 3.2. From graphs to dissimilarities

We focus on the following three main sources from which we can obtain the communities: the node dissimilarity matrices, the eigenvectors of node similarity matrices, and the Euclidean representations of the nodes. Each yields an $n \times n$ dissimilarity matrix $\Delta$ to which any of the linkage criteria above may be applied.

*Node dissimilarity-based community detection.* Hierarchical algorithms can work directly on the distance matrix of the sample. In this approach, we compute the node dissimilarity matrix using one of the measures of Section 2.2 and cluster on it; see Algorithm 1a.

*Eigenvector-based community detection.* In standard spectral community detection, the clustering is performed on the rows of the matrix consisting of the eigenvectors of the graph's Laplacian. To generalise this approach, we substitute the Laplacian with a node similarity matrix $M$ and cluster the rows of a matrix of its eigenvectors under the Euclidean metric; see Algorithm 1b. Following the usual convention for the Laplacian, where the first eigenvector is constant and carries no partition information, we order the eigenvalues of $M$ increasingly and retain the eigenvectors indexed $2, \ldots, K$, so that a $K$-community partition is sought in a $(K-1)$-dimensional space.

*Representation-based community detection.* In the third approach, we first generate vector representations of the nodes with one of the algorithms from Section 2.3 and then apply hierarchical clustering to the Euclidean distances between them; see Algorithm 1c.

*Similarities, dissimilarities and asymmetry.* Several of the measures from Section 2.2 are similarities rather than distances, and must be transformed before use. We use the exponential map $\Delta_{u,v} = \exp(-S_{u,v})$, which is strictly decreasing and therefore preserves the ordering of the proximities on which linkage criteria depend. Two of the measures – the $\ell$-step transition probability of Eq. (8) and Rooted PageRank of Eq. (9) – and the PPMI estimator are not symmetric. We symmetrise them by retaining the upper triangle, i.e., we set $\Delta_{v,u} \leftarrow \Delta_{u,v}$ for $u < v$.

### 3.3. The framework

Algorithm 1 returns the entire dendrogram rather than a single partition. In pipelines a) and c), the dissimilarity matrix does not depend on $K$, so one execution yields every partition at once. In pipeline b), the embedding is $K$-dependent by construction, but the eigendecomposition is computed only once.

*Components.* All three approaches are modular: we may use any appropriate algorithm for calculating the similarity and dissimilarity matrices, the clustering, and the node vector representations. Table 1 lists the components we combine.

### 3.4. Selecting the number of communities

A dendrogram is only useable if a partition can be extracted from it without external information. We therefore treat the choice of $K$ as part of the framework rather than as an input to it. Given the sequence $C^{(1)}, \ldots, C^{(n)}$ produced by Algorithm 1 and an internal criterion $\varphi$, we select:

$$\hat{K} = \arg\max_{K \in \mathcal{K}} \varphi\left(C^{(K)}\right), \tag{11}$$

where $\mathcal{K}$ is a range of admissible cardinalities. We consider two criteria, chosen so that they draw on different information.

- *Modularity* (Definition 2): $\varphi(C) = Q(C)$ evaluated on the input graph. This is the objective that Louvain and Leiden maximise, so the resulting comparison is between methods optimising the same quantity.
- *Silhouette width* [56]: the mean silhouette computed on the dissimilarity matrix $\Delta$ that the pipeline itself produced. This measures whether the geometry induced by the graph model separates the clusters, and makes no reference to modularity.

Both belong to the family of internal stopping rules surveyed by Milligan and Cooper [57]. Modularity inherits the resolution limit [35] discussed in Section 2, and a selection rule based on it can be expected to under-estimate $K$ in the same regimes as the algorithms that optimise

**Algorithm 1** The proposed generic approaches to community detection via hierarchical clustering: a) node dissimilarity-based, b) spectral similarity-based, c) representation-based.

**Input:**
$G$ – the input graph
$K$ – the number of communities (route b only)
**Output:**
$C^{(1)}, \ldots, C^{(n)}$ – a nested sequence of partitions of the nodes of $G$
**Algorithm:**

a)
$\Delta \leftarrow Dissimilarity(G)$

b)
$M \leftarrow Similarity(G)$
$Z \leftarrow$ eigenvectors $2, \ldots, K$ of $M$, eigenvalues increasing
$\Delta \leftarrow$ Matrix of pairwise distances (Euclidean) between all rows in $Z$

c)
$Z \leftarrow GraphRepresentation(G)$
$\Delta \leftarrow$ Matrix of pairwise distances (Euclidean) between all rows in $Z$

$C^{(1)}, \ldots, C^{(n)} \leftarrow HierarchicalClustering(\Delta)$
**return** $C^{(1)}, \ldots, C^{(n)}$

**Table 1**
Framework components tested in our experiments.

| Component type | Tested solutions |
|---|---|
| Similarity and dissimilarity measures (Similarities are converted to dissimilarities by a composition with a strictly decreasing function; see Section 3.2) | Wasserman–Faust measure [13], Neighbourhood overlap (Jaccard) [1], $\ell$-step walk probability [1], Katz index [14], Rooted PageRank [15], Reciprocal-degree index [16], Blondel–Gajardo measure [17], Positive Pointwise Mutual Information [18] |
| Clustering algorithms | Single linkage, complete linkage, average linkage, Ward linkage, Genie [11,12,55] |
| Representation learning algorithms | DeepWalk [19], DNGR [18], Graph Factorisation [20], GraRep [21], HARP [22], HOPE [15], Laplacian Eigenmaps [23], Node2Vec [24] |

it directly. The silhouette carries no such bias but depends on the quality of the induced geometry.

We should note that Eq. (11) maximises the criterion over the $n$ partitions contained in the dendrogram, whereas Louvain and Leiden maximise $Q$ over arbitrary partitions of the graph. When the criterion is modularity, our framework therefore optimises the same objective over a strictly smaller feasible set.

### 3.5. Computational cost

All three routes construct a dense $n \times n$ dissimilarity matrix, so the framework requires $\Theta(n^2)$ memory irrespective of the graph's sparsity. The agglomerative step is $O(n^2 \log n)$ for the linkage criteria we use.

The cost of constructing $\Delta$ depends on the component. For binary $A$, for example, the Wasserman–Faust distance of Eq. (4) admits a closed form that reduces to a single matrix product; the Katz index and Rooted PageRank each require one dense inverse; the PPMI estimator accumulates $T$ successive matrix products. All of these are cubic in $n$. The representation learning methods are dominated by the training procedure of the underlying model rather than by $n$ alone.

## 4. Experimental design

The number of possible (model, parameter, clustering algorithm) triples is large, and reporting the best of many combinations evaluated on the same data would overestimate their performance. This section describes the benchmark data, the parameter grid, the protocol that separates component selection from evaluation, and the statistical methodology employed.

### 4.1. Benchmark data

*Synthetic networks.* The primary benchmark battery consists of graphs generated via the well-known stochastic block model [58]; see Fig. 1 for an illustration.

We generated ten graphs for each combination of the graph parameters presented in Table 2, giving a total of 2880 synthetic graphs ranging from 15 to 200 vertices. Some of these turn out not to be dense enough – particularly those with few vertices per cluster – to guarantee the convergence of all the community detection approaches; the resulting failure rates are reported alongside the results.

This benchmark battery consists of very regular graphs, with communities of equal size and homogeneous densities. It therefore promotes community detection methods that specialise in detecting such network types. This is more or less equivalent to detecting Gaussian blobs in a Euclidean space, where we know that such algorithms as $k$-means and average or complete linkage will outperform the more flexible (nonparametric) methods.

*Real-world networks.* We additionally evaluate the selected pipelines on the six networks listed in Table 3. Zachary's Karate Club [59] and Lusseau's Bottlenose Dolphins [60] are retained to follow the tradition of the existing literature. American College Football [29] and the email-Eu-core network [61,62] are included because their ground truth labels consist of many small communities, a regime that the synthetic battery covers only at its boundary. Cora and CiteSeer [63] are standard citation benchmarks; we note that their labels are topic classes rather than structurally defined communities, so that the attainable agreement is limited for every method [64].

All networks are treated as undirected and unweighted, with self-loops and multiple edges removed. We restrict each graph to its largest connected component, and, where a ground truth is used, discard vertices for which no class label is available; the CiteSeer citation list refers to 17 identifiers that do not occur in its content file. Table 3 reports the vertex and edge counts after this preprocessing.

### 4.2. Components and their parameters

Some of the selected similarity measures have additional parameters that must be set; see Table 2. While the classical hierarchical clustering algorithms do not have any modifiable parameters, for the Genie algorithm, we considered the values suggested in [12]. We assumed the default values for all representation learning algorithms (as in the ReLeGy package for Python [65]) except for the output dimension. The PPMI estimator accumulates $T = 40$ random-surfing steps at every value of the continuation probability.

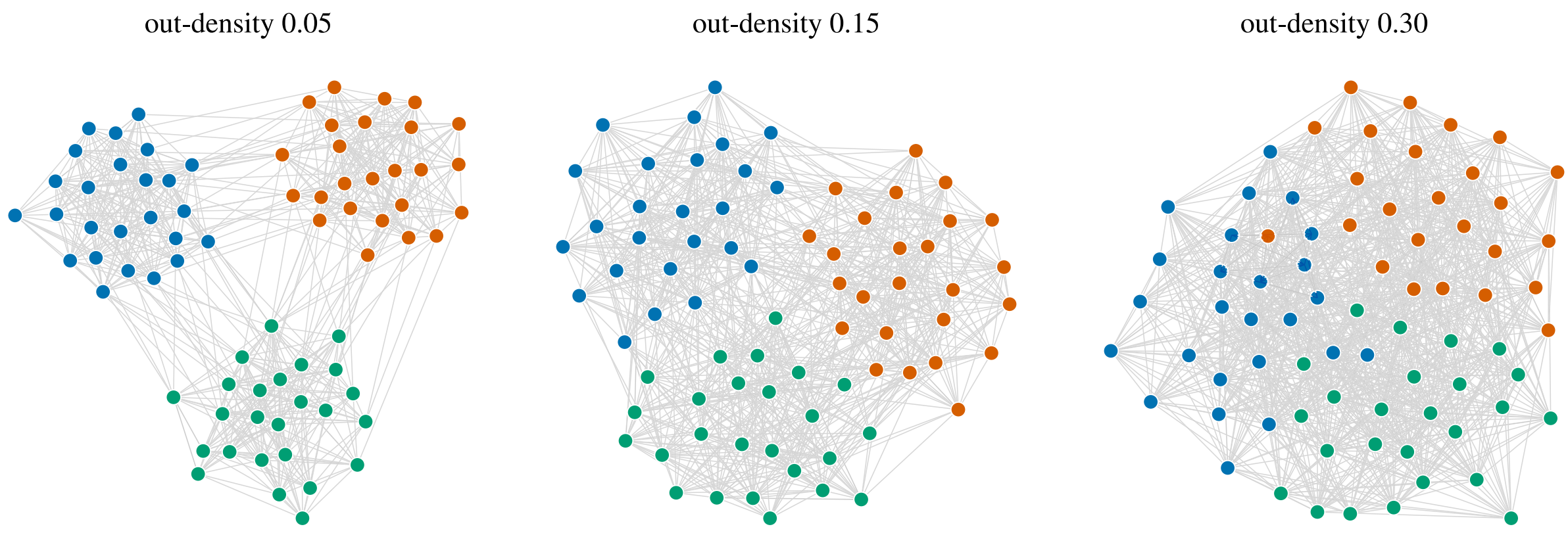

**Fig. 1.** Graphs drawn from the stochastic block model with three communities of 25 vertices and an in-density of 0.6, at three values of the out-density. Colours give the generating block. As the out-density rises, the communities become progressively harder to separate from their surroundings. (For interpretation of the references to colour in this figure legend, the reader is referred to the web version of this article.)

**Table 2**
Parameter grid of all of the tested components.

| Parameter family | Parameter | Values |
|---|---|---|
| Graph | Number of clusters | 3, 4, 5, 10 |
| | Cluster size | 5, 10, 20 |
| | Cluster in-density | 0.4, 0.5, 0.6, 0.7, 0.8, 0.9 |
| | Cluster out-density | 0.05, 0.10, 0.20, 0.30 |
| Component | Length of the $\ell$-step walk | 1, 3, 5 |
| | Katz index decay parameter $\beta$ | 0.1, 0.3, 0.5 |
| | Continuation probability $\alpha$ of Rooted PageRank | 0.5, 0.7, 0.9 |
| | Approximation limit of the Blondel–Gajardo measure | 1, 10, 100 |
| | PPMI continuation probability $\alpha$ ($T = 40$ steps) | 0.7, 0.9, 1.0 |
| Clustering algorithm | Gini index threshold in the Genie algorithm | 0.1, 0.3, 0.5 |
| Representation | Representation dimension | 6, 12 |

**Table 3**
Real-world networks, after restriction to the largest connected component. $K$ denotes the number of ground-truth communities where one is available; for Dolphins, no agreed partition exists so we report modularity instead.

| Network | $\|V\|$ | $\|E\|$ | $K$ | Ground truth |
|---|---|---|---|---|
| Karate | 34 | 78 | 2 | faction |
| Dolphins | 62 | 159 | – | – |
| Football | 115 | 613 | 12 | athletic conference |
| email-Eu-core | 986 | 16,064 | 42 | department |
| Cora | 2485 | 5069 | 7 | topic class |
| CiteSeer | 2110 | 3668 | 6 | topic class |

One grid entry deserves a separate comment. For the Katz index, the series of Eq. (7) converges only when $\beta < 1/\rho(A)$. The spectral radius grows with the density of the generated graphs, so this condition is satisfied across the grid only for the sparsest and smallest members of the battery. For the denser ones, even $\beta = 0.1$ lies outside the radius of convergence of the Neumann series [66], and the matrix that is then computed is no longer a discounted count of walks.

Combining the components of Table 1 with the parameter values of Table 2 gives eighteen parameterised proximity measures (one each for Wasserman–Faust, neighbourhood overlap and the reciprocal-degree index, and three each for the $\ell$-step probability, the Katz index, Rooted PageRank, Blondel–Gajardo and PPMI) and eight representation learning algorithms at two output dimension values. Each is combined with seven clustering settings, namely four linkage criteria and Genie at three Gini index thresholds. The node and spectral routes therefore contribute $18 \times 7 = 126$ configurations each, and the representation route $8 \times 2 \times 7 = 112$, for a total of 364. Of these, 338 returned a result on at least one graph; the remainder are combinations that failed everywhere, chiefly Laplacian Eigenmaps under Genie. One further family was explored and is excluded from this count and from the paper: the Bonferroni index as an alternative inequity measure inside Genie.

### 4.3. Evaluation protocol

*Separating selection from evaluation.* To avoid landing at a biased performance estimator, we split the ten realisations of each parameter combination into two halves. The first five instances form the *screening* set, on which all 364 configurations are compared, and a small number of finalists is chosen. The remaining five cases, 1440 graphs in total, form the *evaluation* set, on which only the finalists and the reference algorithms are run. Each figure that we report for the finalists comes from that half; the experiments of Sections 6.1 and 6.2, which are the most expensive, use three of its five realisations, or 864 graphs.

*Known and unknown $K$.* We report two regimes. In the first, our pipelines are given the true number of communities, which isolates the quality of the graph model from the difficulty of choosing $K$; of the reference algorithms, only standard spectral community detection requires that number; the rest determine it themselves. In the second, the framework selects $K$ from the dendrogram by Eq. (11), and the reference algorithms determine it by their own criteria, so that neither side receives information the other lacks. The search range is $\mathcal{K} = \{2, \ldots, \min(\lfloor n/2 \rfloor, 25)\}$ for the synthetic battery and is widened on the real-world networks whose ground truth exceeds that bound.

*Reference algorithms.* As a point of reference, we use Greedy Modularity Optimisation [67], Label Propagation [68], in the semi-synchronous variant of Cordasco and Gargano [69], the Leiden algorithm [25], the Louvain algorithm [26], and standard spectral community detection [70]. The first four determine the number of communities themselves. Standard spectral community detection, on the other hand, requires $K$ as input, and is therefore supplied with the true value in both regimes. Its results are not directly comparable with the rest, and we mark them accordingly.

*Comparability across methods.* Not every configuration produces a result on every graph, and the failure patterns differ between components. Averages taken over different subsets of the battery are not directly comparable, so all aggregate comparisons and all statistical tests below are computed on the set of graphs solved by every method under consideration.

### 4.4. Performance measures and statistical methodology

Whenever a ground truth partition is available, we measure the agreement using the adjusted Rand index (AR index, see [71]). Assuming that $M$ is a confusion matrix, where $m_{u,v}$ gives the number of elements in the $u$th reference cluster classified by an algorithm as members of the $v$th community,

$$\text{AR-index}(M) = \frac{\binom{n}{2}\sum_{u=1}^{K}\sum_{v=1}^{K}\binom{m_{u,v}}{2} - \sum_{u=1}^{K}\binom{m_{u,\cdot}}{2}\sum_{v=1}^{K}\binom{m_{\cdot,v}}{2}}{\frac{1}{2}\binom{n}{2}\left(\sum_{u=1}^{K}\binom{m_{u,\cdot}}{2} + \sum_{v=1}^{K}\binom{m_{\cdot,v}}{2}\right) - \sum_{u=1}^{K}\binom{m_{u,\cdot}}{2}\sum_{v=1}^{K}\binom{m_{\cdot,v}}{2}}, \tag{12}$$

where $m_{u,\cdot} = \sum_{v=1}^{K} m_{u,v}$ and $m_{\cdot,v} = \sum_{u=1}^{K} m_{u,v}$. The adjusted Rand index attains values between a soft lower limit equal to 0 and a hard upper limit equal to 1 in the case of perfect matches. A partition may yield values slightly lower than 0 if its performance is worse than a purely random assignment. Other measures exist (e.g., [72,73]) and they emphasise different aspects of a partition and need not agree, so we refrain from aggregating them into a single verdict. We report the adjusted Rand index throughout, as is the standard in the literature, and note that the aggregate comparisons below rest on per-graph ranks rather than on absolute index values. Where no ground truth is available, we report modularity given by Eq. (3).

*Statistical comparison.* Aggregate scores are compared with the Friedman test [74] on the per-graph ranks of the methods, followed by, where a specific pair is of interest, the two-sided Wilcoxon signed-rank test [75] on the paired per-graph scores with the Holm correction [76] for multiple comparisons; the procedure follows [77]. With samples of this size, the $p$-value is not by itself informative: differences of a few thousandths of an AR index point attain arbitrarily small $p$-values. We therefore report alongside it the median paired difference, Cliff's effect size measure $\delta$ [78], and, for the ranks, the Nemenyi critical difference at the 5% level, below which two mean ranks are not distinguishable [77].

The median paired difference is the typical gap in adjusted Rand index between two methods on the same graph; a value of zero means that on at least half of the battery the two return partitions of equal quality, however often one is ahead elsewhere. Cliff's $\delta$ is the probability that the first method wins on a randomly chosen graph minus the probability that it loses, so it runs from $-1$ to $+1$ and ignores the size of each win. Values below 0.15 in absolute terms are conventionally read as negligible, 0.15 to 0.33 as small, and 0.33 to 0.47 as moderate [79]. A small $\delta$ with a zero median and a vanishing $p$-value therefore describes a difference that is consistent in direction but slight in magnitude.

**Table 4**
Average (± standard deviation) adjusted Rand index and computation time of the reference community detection algorithms on the 864 evaluation graphs used throughout Section 6. Standard spectral community detection is supplied with the true number of communities; the remaining methods determine it themselves.

| Community detection algorithm | Adjusted Rand index | Time [s] |
|---|---|---|
| Greedy Modularity Optimisation | 0.50 ± 0.35 | 0.03 ± 0.04 |
| Label Propagation | 0.15 ± 0.26 | 0.00 ± 0.00 |
| Leiden | **0.64 ± 0.38** | 0.00 ± 0.00 |
| Louvain | 0.63 ± 0.37 | 0.01 ± 0.02 |
| Spectral | 0.29 ± 0.24 | 0.20 ± 0.22 |

*Run times.* We record the time to compute each graph model. On graphs of the size of the synthetic battery, the time to apply the hierarchical clustering algorithm to an already constructed dissimilarity matrix is negligible; the scaling behaviour of both steps is examined separately in Section 6.4.

## 5. Results on the synthetic datasets

### 5.1. Reference results

Table 4 lists the results for the five reference methods. In line with our expectations, the Leiden and Louvain algorithms achieved the best results, as they are considered state-of-the-art methods for community detection for “regular” graphs such as those generated with the Stochastic Block Model. Greedy Modularity Optimisation was also able to yield good results, albeit clearly worse than the two former methods, while Label Propagation performed rather unsatisfactorily. Standard spectral community detection is not directly comparable with the rest: unlike the other four, it requires the number of communities as an input, and is supplied with the true value.

All of the scores are low relative to the theoretical maximum of 1 because the battery contains configurations that are close to undetectable. Graphs whose communities are barely denser than their surroundings – in-density 0.4 against out-density 0.3, say – defeat every method we tested. The average over the whole battery is therefore not informative on its own, and the comparisons below are made pairwise, on the graphs that every method under consideration solved.

When it comes to the run-time comparison, most of the methods generate the community partitioning almost instantaneously, with the spectral community detection being the only exception; the scaling behaviour of our own pipelines is examined in Section 6.4.

### 5.2. Component screening

Table 5 reports, for each graph model, the parameter value that maximises the mean adjusted Rand index on the screening half of the benchmark battery and the score attained with each clustering algorithm. We note the following three regularities.

First, within the node dissimilarity-based pipeline, the choice of proximity measure matters more than the choice of linkage. PPMI, Wasserman–Faust, the $\ell$-step transition probability, and Rooted PageRank all reach AR of about 0.6, while the Blondel–Gajardo measure does not exceed the level of a random partition at any setting. Average and Ward linkage are the strongest criteria almost throughout, which is consistent with the regularity of the benchmark: communities of equal size in a homogeneous graph are the graph analogue of Gaussian blobs, and those are precisely the structures that the average and Ward linkages recover well. Single linkage is universally subpar.

Second, the spectral pipeline fails to deliver meaningful results. No combination approaches the node dissimilarity-based or representation-based results, and most sit at the level of a random assignment; the best of them, the $\ell$-step transition matrix at 0.196, is less than a third of what the same measure achieves when used directly as a dissimilarity. Section 6.5 takes up the question of why.

**Table 5**
Screening results, measured by the mean adjusted Rand index over five realisations of each scenario. For each graph model, we report the parameter value maximising the mean AR index for each clustering algorithm; the Genie column gives the best result over the Gini index thresholds. "Failed" is the share of the 1440 screening graphs on which the component returned no partition.

| Family | Graph model | Param. | Single | Complete | Average | Ward | Genie | Failed |
|---|---|---|---|---|---|---|---|---|
| Node | PPMI | 1.0 | 0.222 | 0.306 | **0.640** | 0.633 | 0.502 | 4.4% |
| | Wasserman–Faust | – | 0.226 | 0.545 | 0.529 | **0.608** | 0.521 | 0.0% |
| | $\ell$-step walk probability | 3 | 0.240 | 0.523 | 0.546 | **0.608** | 0.476 | 4.4% |
| | Rooted PageRank | 0.7 | 0.179 | 0.527 | **0.579** | 0.577 | 0.449 | 0.0% |
| | Reciprocal-degree index | – | 0.158 | 0.365 | 0.459 | **0.536** | – | 4.4% |
| | Neighbourhood overlap (Jaccard) | – | 0.259 | 0.352 | 0.508 | **0.521** | 0.444 | 0.6% |
| | Katz index | 0.1 | 0.074 | 0.347 | 0.335 | **0.429** | 0.298 | 0.0% |
| | Blondel–Gajardo | 1 | 0.001 | 0.003 | 0.004 | 0.007 | **0.018** | 0.0% |
| Spectral | $\ell$-step walk probability | 5 | 0.014 | 0.090 | 0.033 | 0.158 | **0.196** | 4.4% |
| | Katz index | 0.3 | 0.130 | 0.174 | 0.176 | 0.187 | **0.189** | 0.0% |
| | Wasserman–Faust | – | 0.003 | −0.001 | 0.001 | 0.001 | **0.043** | 4.4% |
| | Rooted PageRank | 0.9 | 0.004 | 0.012 | 0.006 | 0.021 | **0.030** | 0.0% |
| | Blondel–Gajardo | 1 | 0.005 | 0.018 | 0.009 | 0.023 | **0.025** | 0.0% |
| | PPMI | 1.0 | – | −0.003 | 0.001 | −0.004 | **0.014** | 4.4% |
| | Neighbourhood overlap (Jaccard) | – | 0.001 | −0.008 | −0.001 | −0.010 | **0.005** | 4.4% |
| | Reciprocal-degree index | – | −0.001 | −0.019 | −0.008 | −0.023 | −0.008 | 4.4% |
| Representation | DNGR | 6 | 0.392 | 0.541 | 0.539 | **0.595** | 0.575 | 4.4% |
| | Node2Vec | 6 | 0.209 | 0.459 | 0.460 | **0.524** | 0.481 | 4.4% |
| | DeepWalk | 6 | 0.206 | 0.448 | 0.432 | **0.521** | 0.476 | 0.0% |
| | GraRep | 6 | 0.133 | 0.310 | 0.246 | **0.448** | 0.420 | 4.4% |
| | HARP | 6 | 0.176 | 0.386 | 0.358 | **0.446** | 0.411 | 0.0% |
| | Graph Factorisation | 12 | 0.162 | 0.342 | 0.281 | 0.410 | **0.411** | 0.0% |
| | HOPE | 12 | 0.046 | 0.146 | 0.132 | 0.296 | **0.302** | 0.0% |
| | Laplacian Eigenmaps | 6 | 0.020 | 0.060 | 0.043 | **0.085** | – | 68.7% |

Third, among the representation learning methods, DNGR is the strongest by a wide margin, followed by the two random-walk methods, Node2Vec and DeepWalk. The factorisation-based methods trail them, and Laplacian Eigenmaps does not produce useable partitions. Almost all of the well-performing methods prefer the six-dimensional representation, which is unsurprising given the size of the test graphs; larger graphs would be expected to require more dimensions.

*Scope of the search.* Combining the components of Table 1 with the parameter values of Table 2 gives the 364 configurations enumerated in Section 4.2, all of which were screened, and of which 338 returned a result on at least one graph.

*Numerical failures.* Not every configuration returns a partition on every graph, and Table 5 reports the share of the screening half on which each component failed. Most components lose about 4.4% of the battery, always on the same graphs: the sparsest and smallest configurations, where a generated graph can be disconnected or contain isolated vertices. Laplacian Eigenmaps is a different case, returning a result on only 451 of the 1440 screening graphs in six dimensions and, for two of the four linkage criteria, on fewer than one in ten. Its scores are reported for completeness but rest on too small and too selected a subsample to be compared with the others.

### 5.3. Selected pipelines

On the shoulders of the conclusions arising from analysing data in Table 5, we carry the following nine cases forward. The remainder of the paper evaluates them on the held-out benchmark instances, which took no part in the selection.

- *Node Genie PPMI* – node dissimilarity-based Genie with PPMI-based dissimilarity, the Gini index threshold equal to 0.1, and continuation probability equal to 1,
- *Node Genie W-F* – node dissimilarity-based Genie with Wasserman–Faust measure-based dissimilarity, and the Gini index threshold equal to 0.1,
- *Node Average PPMI* – node dissimilarity-based average linkage with PPMI-based dissimilarity, and continuation probability equal to 1,
- *Node Average RPR* – node dissimilarity-based average linkage with Rooted PageRank-based dissimilarity, and continuation probability equal to 0.7,
- *Spectral Genie $\ell$-step* – spectral-based Genie with $\ell$-step probability-based similarity, threshold equal to 0.1, and number of steps equal to 5,
- *Spectral Ward Katz* – spectral-based Ward linkage with Katz index-based similarity, and decay parameter equal to 0.3,
- *Euclidean Genie DNGR* – representation-based Genie with DNGR representation, threshold equal to 0.1, and embedding dimension equal to 6,
- *Euclidean Ward DNGR* – representation-based Ward linkage with DNGR representation, and embedding dimension equal to 6,
- *Euclidean Ward Node2Vec* – representation-based Ward linkage with Node2Vec representation, and embedding dimension equal to 6.

The two spectral combinations are retained despite their weak screening scores: they are the only members of our pipeline worth examining, and their behaviour turns out to be diagnostic of why the pipeline as a whole does not work.

## 6. Detailed evaluation

Let us compare the nine selected methods against the five reference methods listed in Section 5.1.

### 6.1. Detection with a known number of communities

We first compare the selected pipelines with the reference algorithms in the setting used throughout the earlier literature, in which our pipelines are supplied with the true number of communities. Of the reference algorithms, only standard spectral community detection requires that number; Leiden, Louvain, Greedy Modularity Optimisation and Label Propagation determine it themselves.

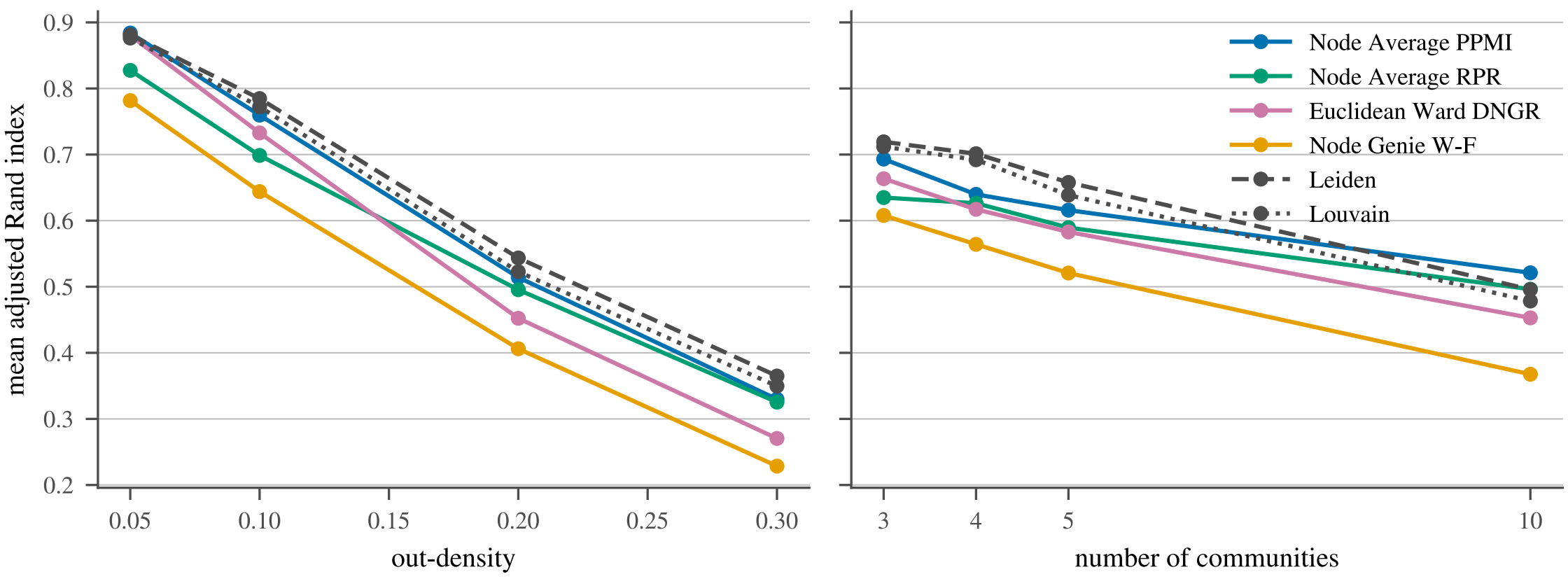

**Fig. 2.** Mean adjusted Rand index against the out-density of the communities (left) and their number (right), for the four strongest pipelines and the two strongest reference algorithms. Our pipelines are given the true number of communities; Leiden and Louvain determine it themselves.

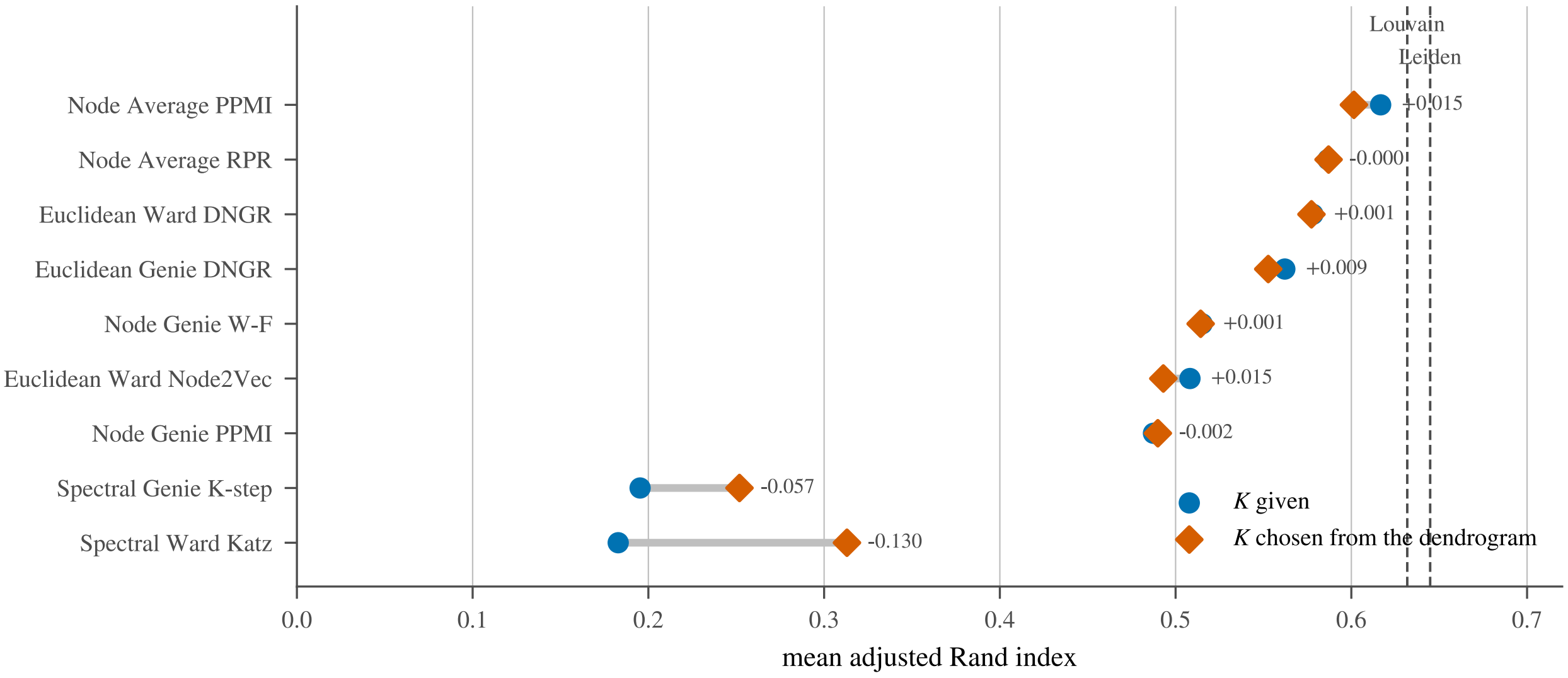

**Fig. 3.** The cost of estimating the number of communities from the dendrogram rather than having it given a priori. Circles mark the mean adjusted Rand index at $K = K_{\text{true}}$, diamonds designate the score at the cardinality selected by the internal criterion, and the figure at the right of each pair is the difference. Dashed lines give the reference algorithms, which determine the number of communities themselves.

Fig. 2 summarises the outcome along the two axes that govern the difficulty of the benchmark. Each node dissimilarity- and representation-based pipeline behaves as a community detection method should: performance falls steadily as the out-density rises and the communities become less distinguishable from their surroundings, from around 0.88 at an out-density of 0.05 to below 0.35 at 0.30. Performance also declines as the communities become more numerous, since a fixed budget of vertices divided among more groups leaves each of them smaller and the sparse regions between them narrower.

Node Average PPMI is the strongest of our pipelines, at 0.617 over the whole battery against 0.644 for Leiden and 0.630 for Louvain, and it tracks the two references closely across the density range. The right panel of Fig. 2 shows the one place where the ordering changes: at ten communities our pipeline reaches 0.521 against 0.496 for Leiden and 0.478 for Louvain. Section 6.2 takes up that crossing, which turns out to be the more informative part of the comparison.

The two spectral pipelines are absent from Fig. 2 because they do not reach a level at which these comparisons are meaningful, and because they behave unlike every other method in the study; Section 6.5 examines them separately.

### *6.2. Detection without a known number of communities*

The property that distinguishes our framework from the reference algorithms is that a single run yields a partition at every cardinality. That is only useful if a good cardinality can be recovered from the dendrogram without being told what it is. This section measures what that costs.

*The price of not knowing $K$.* Fig. 3 compares, for each pipeline, the score obtained when the true number of communities is supplied against the score obtained when $K$ is chosen from the dendrogram via Eq. (11). For the seven non-spectral pipelines, the AR difference lies between −0.003 and 0.015: choosing the cardinality from an internal criterion recovers at least 97% of the performance available with oracle knowledge. Two pipelines lose less than a thousandth of a point, and for two others – Node Genie PPMI and Node Average RPR – the selected partition is marginally better than the one at $K = K_{\text{true}}$.

Two things could have gone wrong, and neither did. The dendrogram might have contained no partition as accurate as the one at $K_{true}$ at any other cardinality, so that any error in $\hat{K}$ would be paid for in adjusted Rand index; that would be a defect of the hierarchy. Or such a partition might have been present, and the internal criterion have cut elsewhere; that would be a defect of the selection rule. The differences in Fig. 3 are too small for either.

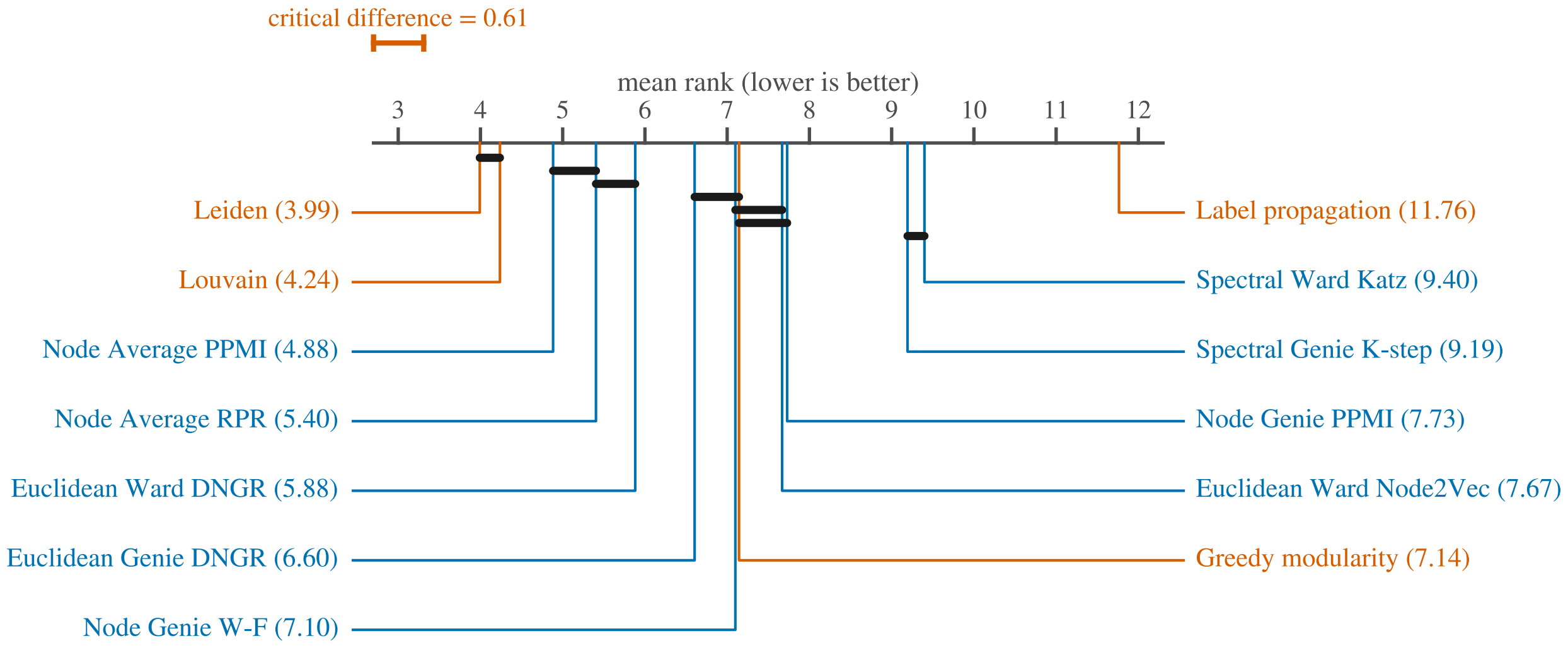

**Fig. 4.** Critical-difference diagram of mean ranks over the 844 graphs solved by every method, with the number of communities discovered on both sides. Methods joined by a bold horizontal bar are not distinguishable from one another at the 5% significance level. Reference algorithms are shown in orange. (For interpretation of the references to colour in this figure legend, the reader is referred to the web version of this article.)

*A like-for-like comparison.* With $K$ discovered on both sides, Fig. 4 ranks the nine pipelines against the four reference algorithms that also determine it themselves, on the 844 graphs on which neither algorithm failed. The Friedman test rejects the hypothesis of equal performance ($\chi^2_{12} = 4064$, $p < 10^{-16}$). Leiden ranks first at 3.99 and Louvain second at 4.24; the two are not distinguishable from each other at the Nemenyi critical difference of 0.61. Node Average PPMI follows at 4.88 and is distinguishable from both, though the margin against Louvain (0.65) exceeds the critical difference only narrowly. Below them, Node Average PPMI is not separable from Node Average RPR, nor Node Average RPR from Euclidean Ward DNGR, although the span from the first of the three to the last does exceed the critical difference. All three are separated from Greedy Modularity Optimisation in the other direction.

The size of the gap is better read from the paired comparisons than from the ranks. Table 6 gives the two-sided Wilcoxon signed-rank tests of Node Average PPMI against each reference algorithm, with $p$-values Holm-corrected over the four comparisons. Every difference is significant at any conventional level, which at this sample size is to be expected and is not by itself informative. The effect sizes are: against Leiden a mean difference of −0.043 with Cliff's $\delta = -0.21$, against Louvain −0.030 with $\delta = -0.14$, both small by the convention of Section 4.4. In both cases the median paired difference is exactly zero, and the two methods return partitions of identical quality on about 29% of the battery. In the other direction, the margins are substantial: +0.100 against Greedy Modularity Optimisation ($\delta = +0.40$) and +0.462 against Label Propagation ($\delta = +0.91$).

*Where the framework wins, and why.* The aggregate hides a systematic pattern. The left panel of Fig. 5 gives the difference in adjusted Rand index between Node Average PPMI and Leiden across the cluster structures in the battery. Leiden is ahead almost everywhere, but in the regime of many small communities the ordering reverses: at ten communities of five vertices our pipeline gains 0.115 (Wilcoxon, $\delta = +0.61$, Holm-corrected $p < 10^{-8}$). It is the only cell in which either method is significantly ahead of the other in our favour; at ten communities of ten vertices the two are indistinguishable (+0.017, $p = 0.35$), and Leiden leads elsewhere.

**Table 6**
Node Average PPMI against the reference algorithms on the 844 graphs solved by all methods, with the number of communities discovered on both sides. Two-sided Wilcoxon signed-rank tests, $p$-values Holm-corrected over the four comparisons.

| Reference | Mean diff. | Median diff. | Cliff's $\delta$ | Holm $p$ |
|---|---|---|---|---|
| Leiden | −0.043 | 0.000 | −0.21 | $1.9 \times 10^{-15}$ |
| Louvain | −0.030 | 0.000 | −0.14 | $7.7 \times 10^{-8}$ |
| Greedy Modularity Optimisation | +0.100 | +0.048 | +0.40 | $3.4 \times 10^{-41}$ |
| Label Propagation | +0.462 | +0.414 | +0.91 | $1.6 \times 10^{-130}$ |

The right panel explains it. It reports how many communities Leiden fails to separate, $K_{true}$ minus the number it returns. At ten communities of five vertices, Leiden recovers 5.6 of the 10 present; the deficit falls to 2.7 at ten vertices per community and to 1.4 at twenty, and reverses sign for three communities, where Leiden slightly over-partitions. This is the resolution limit of modularity described in Section 2, visible directly: modularity maximisation merges communities joined by too few edges, and the regime in which it does so is exactly the regime where our framework is ahead. The mechanism and the effect are measured on the same graphs.

*Which criterion to cut on.* The modularity rule of Eq. (11) is the better default, outperforming the silhouette for eight of the nine pipelines and recovering the true cardinality more often (exact on 42%–54% of graphs against Leiden's 61%, with a mean absolute error between 1.0 and 1.7 communities against Leiden's 0.96). The exception is Node Average PPMI, for which the silhouette is marginally better. A modularity-based cut inherits the same resolution limit as the algorithms that maximise modularity directly, which is a liability in precisely the regime where our framework is otherwise strongest.

### 6.3. Real-world networks

Fig. 6 reports the agreement with the ground truth on the five labelled networks of Table 3, with every method choosing the number of communities itself. To avoid reintroducing the selection problem, we nominate a single primary pipeline – Node Average PPMI under the silhouette rule, the best-performing configuration on the synthetic evaluation – as a representative of our framework on every network. The other pipelines are shown in the figure for reference, and we note where one of them performs better.

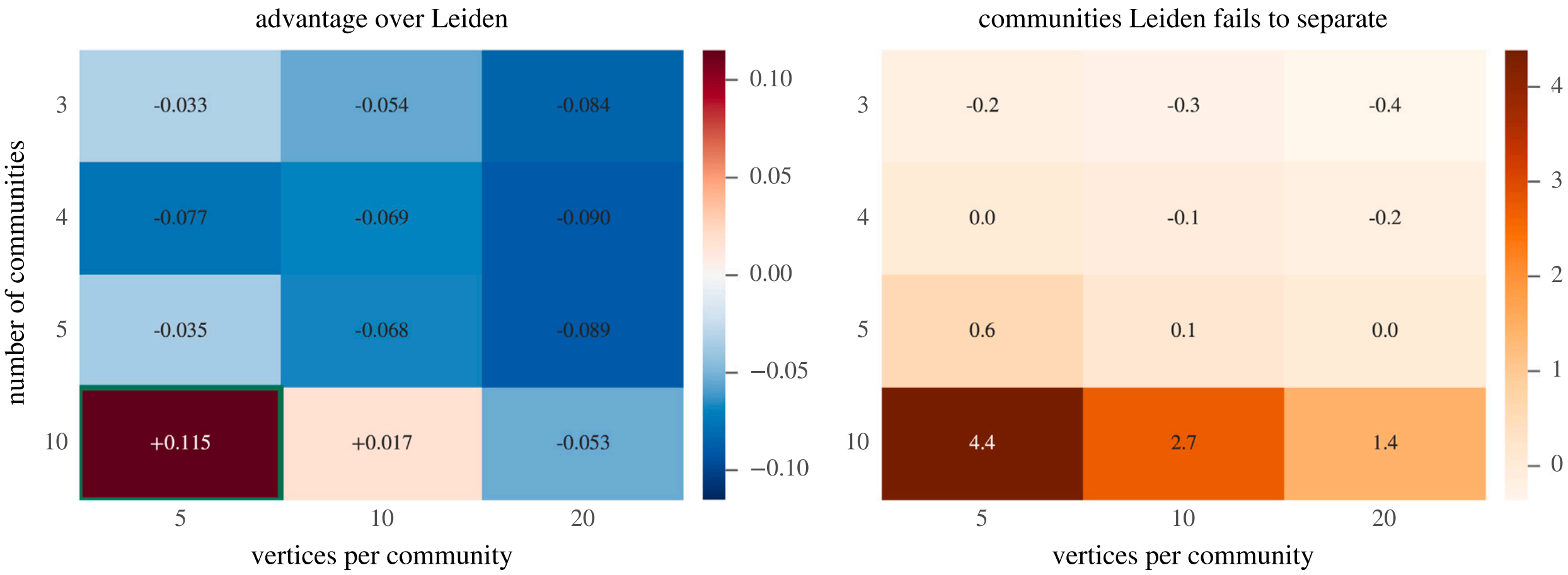

**Fig. 5.** Left: the difference in adjusted Rand index between Node Average PPMI under its silhouette rule and Leiden, both selecting the number of communities themselves; the green outline marks the one cell where our pipeline is ahead at the 5% significance level according to the Wilcoxon signed-rank test Holm-corrected over the twelve cells. Right: the average number of communities Leiden fails to separate, $K_{\text{true}}$ minus the number it returns. The framework is ahead in the cell where modularity maximisation under-partitions most severely.

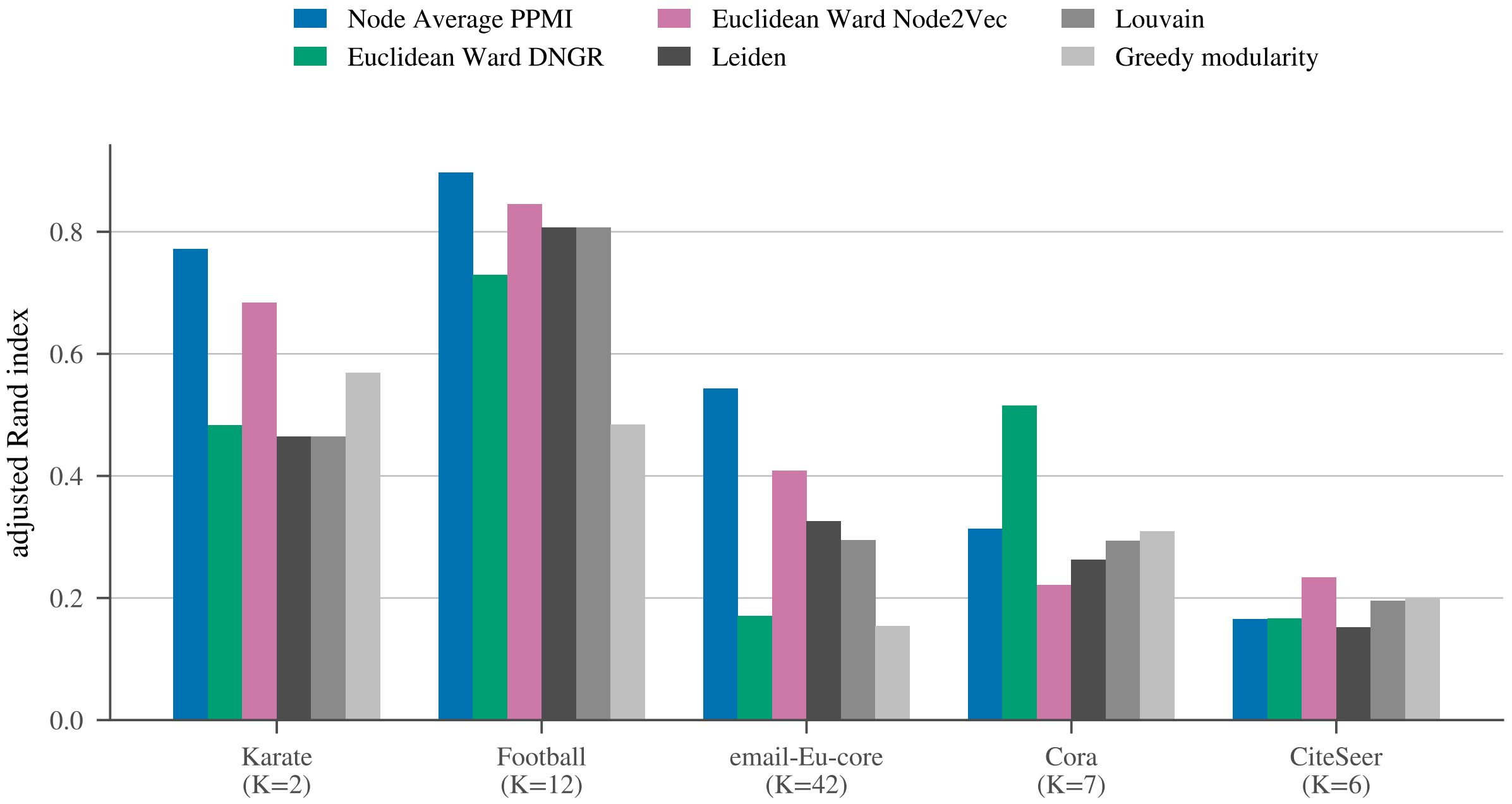

**Fig. 6.** Agreement with the ground truth on the five labelled real-world networks, with every method determining the number of communities itself. Each pipeline uses the selection criterion established in Section 6.2 – modularity for all but Node Average PPMI, which uses the silhouette – across all five networks.

On Karate, our primary pipeline reaches AR of 0.772 and recovers the reference split at the correct cardinality, versus AR of 0.465 for Leiden and Louvain, both of which return four communities where the ground truth has two. Greedy Modularity Optimisation, at three communities, reaches AR of 0.568. Note that this is *over*-partitioning??, i.e., detecting $\hat{K} > K_{true}$ communities, rather than the resolution limit of Section 6.2, which incorrectly merges real communities, often resulting in $\hat{K} > K_{true}$; the right panel of Fig. 5 shows the same over-partitioning by Leiden at small $K$ on synthetic graphs. What Karate demonstrates is, therefore, the value of selecting the cardinality from a dendrogram, not the resolution-limit mechanism.

On email-Eu-core, whose 42 departments make it the most fragmented ground truth we consider, the primary pipeline reaches AR of 0.543 versus 0.326 for Leiden and 0.294 for Louvain. Both references settle on eight communities, a fifth of the number present. On Football, it recovers the twelve athletic conferences at the correct cardinality, scoring AR of 0.897 versus 0.807 for Leiden and Louvain, which return ten. These two networks occupy the regime where the framework is ahead on synthetic data, namely many communities with few vertices in each, and the resolution-limit mechanism of Section 6.2 applies to both.

The two citation networks behave differently. Their labels are topic classes assigned to documents, not groups defined by the link structure, so no method can be expected to recover them from topology alone [64]; the attainable agreement is low for everything we ran. Within that ceiling, our primary pipeline reaches AR of 0.305 on Cora and 0.102 on CiteSeer, versus 0.309 and 0.200 for the best reference; it is competitive on the first and lags on the second. A representation-based pipeline does substantially better on Cora, with Euclidean Ward DNGR reaching 0.515. Note also that Leiden and Louvain partition these networks into 29 to 37 communities against a ground truth of six or seven, and that our pipeline does the same, selecting 32 and 58.

For Dolphins, where no agreed partition exists, we report modularity instead: Leiden attains 0.528, Euclidean Genie DNGR 0.519

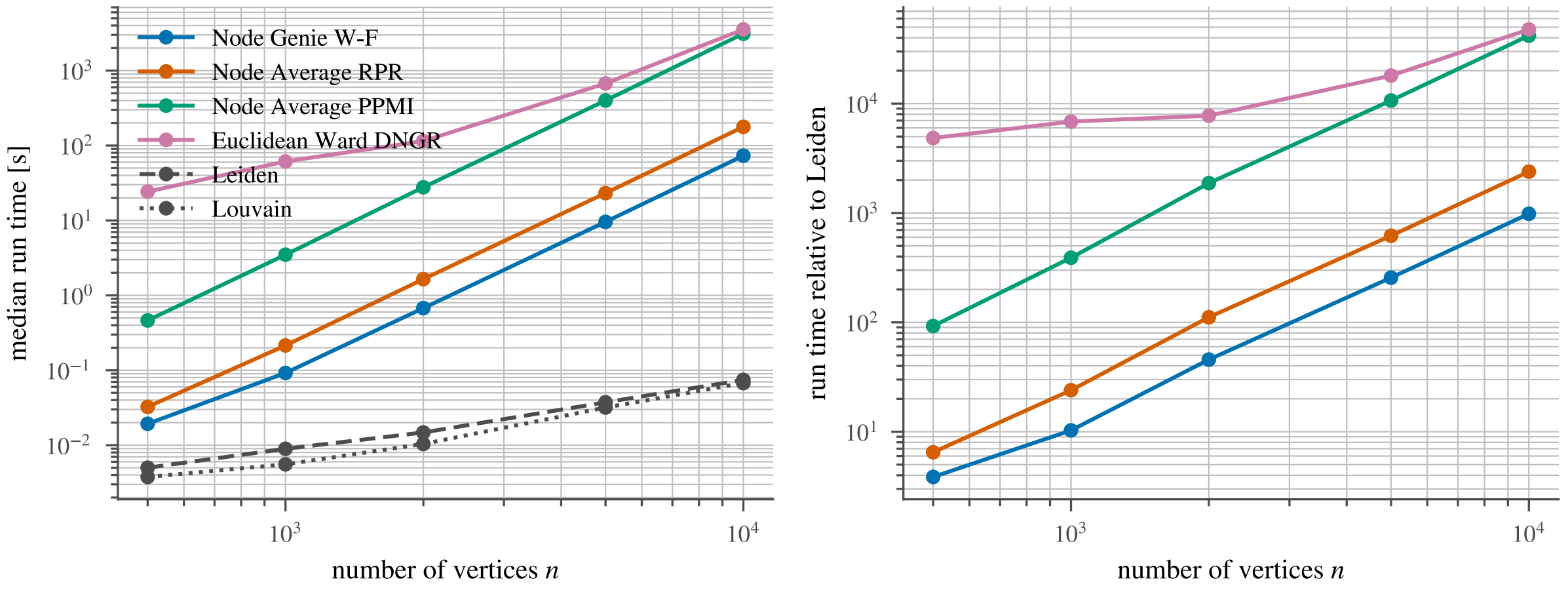

**Fig. 7.** Median run time as a function of graph size in the sparse regime, absolute (left) and relative to Leiden (right), on a log scale.

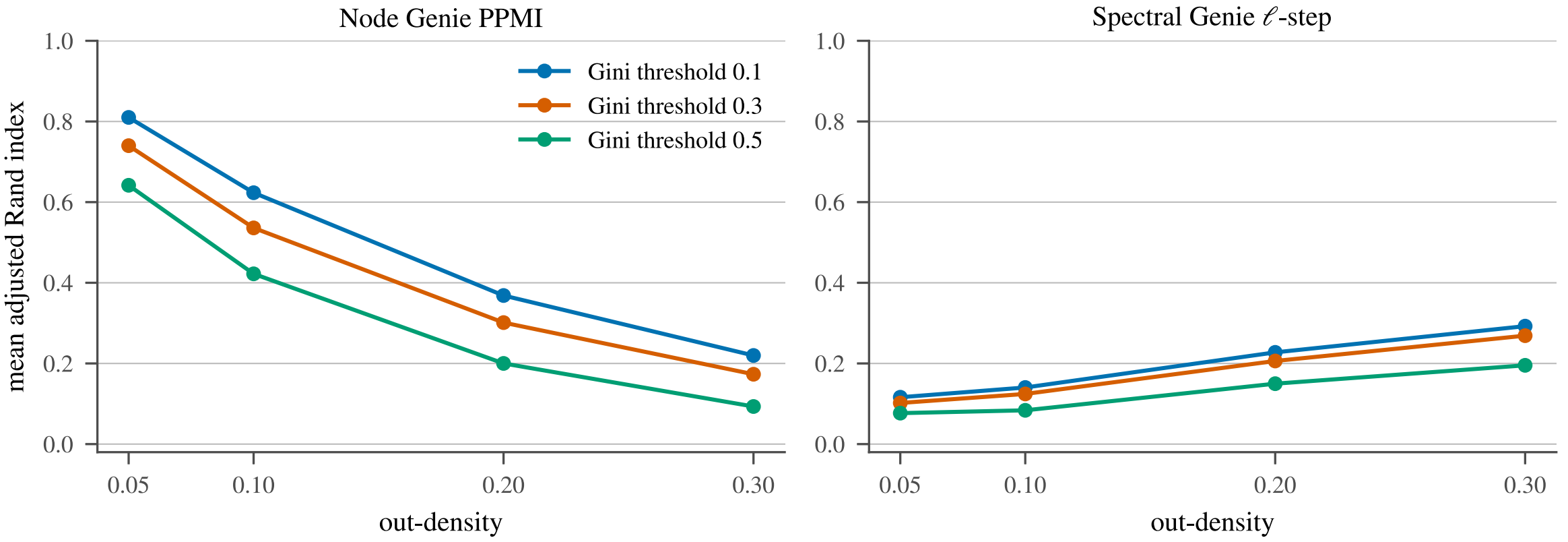

**Fig. 8.** Mean adjusted Rand index against the out-density of the communities, at each Gini index threshold. A pipeline that detects communities improves as they become easier to separate (left); the spectral pipeline does the opposite (right).

and Louvain 0.519. Our framework is competitive but not ahead of them, which is the expected outcome on a criterion that the reference algorithms optimise directly.

### *6.4. Scalability*

The synthetic battery caps at 200 vertices, which is too small to say anything about the computational cost. We therefore generated a separate series of stochastic block model graphs in a sparse regime, with communities of 100 vertices and edge probabilities scaled so that the expected degree stays near 16 as the graph grows; this holds the strength of the community signal fixed while $n$ varies from 500 to 10,000.

Fig. 7 reports the results. Run times are medians over three realisations on a single thread. Leiden and Louvain scale near-linearly. Our pipelines scale as $O(n^3)$ as the $\ell$-step and PPMI constructions accumulate matrix products, and the Rooted PageRank inverts a dense matrix.

We also observe that the agglomerative step is not the bottleneck: at $n = 10{,}000$ the linkage takes one to two seconds against the 3548 s Euclidean Ward DNGR spends building its representation, so the cost lies entirely in constructing the graph model. The framework is therefore suited to networks up to roughly $10^4$ vertices on a typical contemporary personal computer.

### *6.5. Why the spectral pipeline fails*

The spectral pipeline was included to test whether the standard spectral construction [80] generalises when the Laplacian is replaced by an arbitrary vertex similarity matrix. Our results point towards a negative answer.

Part of the reason is algebraic. The Katz matrix in Eq. (7) is an analytic function of the adjacency matrix [66], $S^{\mathrm{Katz}} = \beta(I - \beta A)^{-1}A$, so it shares its eigenvectors with $A$ and its eigenvalues are the images $\beta\lambda/(1-\beta\lambda)$ of the eigenvalues of $A$. Two cases must be distinguished, and in both this method gains nothing.

When $\beta\rho(A) < 1$, the map $\lambda \mapsto \beta\lambda/(1-\beta\lambda)$ is increasing on the whole spectrum, so ordering by eigenvalue selects the same subspace as ordering the eigenvectors of $A$ itself: spectral clustering on the Katz matrix is spectral clustering on the adjacency matrix in disguise. When $\beta\rho(A) > 1$, which as Section 4.2 notes covers most of our battery, the map is undefined at $\lambda = 1/\beta$ and is no longer monotone across the spectrum; the eigenvalue ordering is then permuted in a way that depends on where the denominator zero falls, and the matrix is in any case not a discounted count of walks. The first case adds nothing to the standard construction and the second has no interpretation in terms of graph structure. The same dichotomy applies to Rooted PageRank as a function of the transition matrix $P$, with the difference that $\rho(P) = 1$ places the whole grid of Table 2 inside the convergent case. This accounts for the scores of both in Table 5 sitting close to those of standard spectral community detection.

The $\ell$-step matrix is the one combination that behaves differently, and Fig. 8 shows why its apparent success should not be trusted. A pipeline that reads community structure must do better when the communities are easier to see: the mean adjusted Rand index of Node Genie PPMI falls from 0.81 to 0.22 as the out-density rises from 0.05 to 0.30. Spectral Genie $\ell$-step does the reverse, rising from 0.12 to 0.29, and it improves monotonically as the Gini threshold is tightened. Since every graph in the battery has communities of equal size, a partition forced towards equal cardinalities is rewarded wherever the graph carries no useable signal, and that is what the rising curve records.

A second observation points the same way. Both spectral pipelines score higher in terms of mean AR-index when the number of communities is chosen from the dendrogram than when the true number is supplied: 0.313 versus 0.183 for Spectral Ward Katz, and 0.252 versus 0.195 for Spectral Genie $\ell$-step (Fig. 3). A method that tracks the community structure of a graph cannot improve by being denied the correct number of communities.

We therefore report the spectral pipeline as a negative result. Spectral clustering works on the Laplacian because its leading eigenvectors identify splits of the graph that cut few edges; the eigenvectors of an arbitrary similarity matrix carry no such guarantee, and the one combination that appeared promising is an artefact of the benchmark's regularity.

## 7. Discussion and conclusions

We set out to determine whether hierarchical clustering algorithms, applied to graphs through a vertex proximity matrix, can compete with dedicated community detection methods, and what such an approach costs. Three general pipelines were developed for the purpose – node dissimilarity-based, spectral, and representation-based – and 364 configurations of their components were compared on 2880 stochastic block model graphs and six real-world networks. The following is what we take from this exercise.

*A dendrogram is nearly free.* The property that distinguishes this framework from Leiden and Louvain is that one run yields a partition at every cardinality. The natural objection is that this is bought at the price of accuracy, since a partition must still be chosen from the hierarchy. Section 6.2 shows the price to be between −0.003 and 0.015 of the adjusted Rand index: cutting the dendrogram by an internal criterion recovers at least 97% of what is available when the true number of communities is supplied. The framework's central claim is therefore not a trade-off in any practical sense.

*Our framework is competitive.* With the number of communities discovered on both sides, Node Average PPMI ranks third of thirteen methods, behind Leiden and Louvain and ahead of Greedy Modularity Optimisation and Label Propagation. The gap to the two leaders is consistent in direction but slight in size, with a median paired difference of zero. Against that, the ordering reverses in a regime we can characterise precisely: when a graph contains many small communities, our pipelines are ahead, and the margin grows as the communities get smaller. Section 6.2 attributes this to the resolution limit of modularity, and measures the mechanism directly: at ten communities out of five vertices, Leiden recovers fewer than six of the ten present, and the deficit shrinks in step with our advantage. The effect reproduces on real networks with many communities, most clearly on email-Eu-core, where 42 departments are met with eight communities by both references.

*Representation learning is effective, but costly.* Of the eight representation learning methods, DNGR is the strongest and reaches the level of the best proximity measures on the synthetic battery, while being the most expensive component we tested by a wide margin. On the citation networks, it is the best method overall. The remaining methods do not justify their cost here, and two of them do not produce useable partitions. Whether a learned representation is worth its price appears to depend on the network rather than being settled in general.

*The spectral generalisation does not work.* Substituting an arbitrary similarity matrix for the Laplacian removes the property that makes spectral clustering effective, and for two of the measures we show the substitution to be disadvantageous. We report this as a negative result.

*When to use this framework.* The picture that emerges is a clear division of cases. On networks up to roughly $10^4$ vertices, where the number of communities is unknown or expected to be large relative to the size of the graph, and where nested partitions at every cardinality are wanted, the framework is a reasonable choice and will sometimes be the better one. Outside those conditions, Leiden and Louvain remain preferable, and the run-time gap that Section 6.4 documents grows by an order of magnitude for every decade of graph size.

*Limitations and future work.* The synthetic evaluation rests on a single generative model whose communities are of equal size, which favours methods that produce balanced partitions; the benchmarks of Lancichinetti et al. [81] would test the framework against heterogeneous community sizes and degree distributions, and we regard that as the most useful next step. The framework consumes only topology, so attributed, temporal and overlapping community detection lie outside its present scope. Moreover, the presented methods do not translate cleanly to directed networks, such as citation networks with communities [82]. Finally, the $\Theta(n^2)$ cost is intrinsic to working with a dense proximity matrix; a sparse or approximate construction would be required before the approach could be applied at the scale at which Leiden and Louvain are routinely used.

In summary, we have shown that agglomerative hierarchical clustering, applied to graphs through a vertex proximity matrix, is a viable approach to community detection. Across 2880 synthetic graphs and six real-world networks, the strongest of our pipelines ranks immediately behind Leiden and Louvain when all three determine the number of communities for themselves, and ahead of them on networks composed of many small communities, where modularity maximisation is known to under-partition and where we measure it doing so.

The property that motivated the approach survives quantification. Selecting a partition from the dendrogram costs almost nothing relative to being told the correct number of communities, which means the hierarchy is available at no practical loss. Set against that, our framework is slower than Leiden and does not extend far beyond the scale of around $10^4$ vertices.

Of the three pipelines proposed, two are potentially useful: the node dissimilarity-based approach, where positive pointwise mutual information and the Wasserman–Faust measure are the strongest models, and the representation-based approach, where DNGR is. The spectral pipeline does not work, and we have explained why. We regard the resulting picture – a thorough benchmark yielding a competitive method in general and better than the state of the art in a regime that can be stated in advance – as a useful addition to the literature on hierarchical clustering and community detection.

## CRediT authorship contribution statement

**Łukasz Brzozowski:** Writing – review & editing, Writing – original draft, Visualization, Software, Methodology, Data curation, Conceptualization. **Grzegorz Siudem:** Writing – review & editing, Methodology. **Marek Gagolewski:** Writing – review & editing, Methodology, Conceptualization.

## Declaration of competing interest

All authors certify that they have no affiliations with or involvement in any organisation or entity with any financial interest or non-financial interest in the subject matter or materials discussed in this manuscript.

## Acknowledgements

The research was carried out with the support of the Laboratory of Bioinformatics and Computational Genomics and the High Performance Computing Centre of the Faculty of Mathematics and Information Science, Warsaw University of Technology under computational grant number A-21-04. The first author would like to thank Kacper Siemaszko and Paweł Rzążewski for stimulating discussion.

## Data availability

The real-world networks used in this study – Zachary's Karate Club, Lusseau's Bottlenose Dolphins, American College Football, email-Eu-core, Cora and CiteSeer – are freely available online. Our battery of synthetic stochastic block model graphs, and the code implementing the framework, will be made available upon request.